\documentclass[twocolumn,twoside,slac]{revtex4}
\usepackage{graphicx}
\usepackage{fancyhdr}
\def\babar{\mbox{\sl B\hspace{-0.4em} {\small\sl A}\hspace{-0.37em} \sl B\hspac
e{-0.4em} {\small\sl A\hspace{-0.02em}R}}}
\usepackage{relsize} \def\babar{\mbox{\slshape B\kern-0.1em{\smaller A}\kern-0.1em B\kern-0.1em{\smaller A\kern-0.2em R}}}

\begin{document}

%Title of paper
\title{The \babar{} Event Building and Level-3 Trigger Farm Upgrade}

\author{R. Jacobsen}
\affiliation{Lawrence Berkeley National Laboratory and University of 
California, Berkeley, California 94720}
\author{G. Dubois-Felsmann, I. Narsky}
\affiliation{California Institute of Technology, Pasadena, California 
91125}
\author{E. Rosenberg}
\affiliation{Iowa State University, Ames, Iowa 50011-3160}
\author{B. Franek}
\affiliation{Rutherford Appleton Laboratory, Chilton, Didcot, Oxon, OX11 
0QX United Kingdom}
\author{R. Bartoldus, J. Hamilton, D. Kotturi, S. Luitz, C. O'Grady,
         A. Perazzo, R. Rodriguez, A. Salnikov, M. Weaver, M. Wittgen}
\affiliation{SLAC, Stanford, CA 94025, USA}
\author{S. Dasu}
\affiliation{University of Wisconsin, Madison, Wisconsin 53706}

\author{\vspace{1em}for the \babar{} Computing Group}

\begin{abstract}
The \babar{} experiment is the particle detector at the PEP-II B-factory
facility at the Stanford Linear Accelerator Center \cite{babardetector}.
During the summer shutdown 2002 the \babar{} Event Building and Level-3
trigger farm were upgraded from 60 Sun Ultra-5 machines and 100MBit/s
Ethernet to 50 Dual-CPU 1.4GHz Pentium-III systems with Gigabit
Ethernet. Combined with an upgrade to Gigabit Ethernet on the source
side and a major feature extraction software speedup, this pushes the
performance of the \babar{} event builder and L3 filter to 5.5kHz at
current background levels, almost three times the original design rate
of 2kHz. For our specific application the new farm provides 8.5 times
the CPU power of the old system.

\end{abstract}

\maketitle

\thispagestyle{fancy}

\section{THE \babar{} DATA ACQUISITION SYSTEM}

The primary data handling components of the \babar{} data acquisition
system are the {\em Online Data Flow} (ODF) \cite{babarodf} and the
{\em Online Event Processing} (OEP) \cite{babaroep}. ODF is the
real-time system for trigger distribution, readout of the front end
electronics, feature extraction and multi-stage event building. OEP
provides near-real-time processing of fully built events: Level-3
software trigger, fast data quality monitoring, calibration processing
and dispatch to data logging.

\begin{figure*}[t]
\centering
\includegraphics[width=135mm]{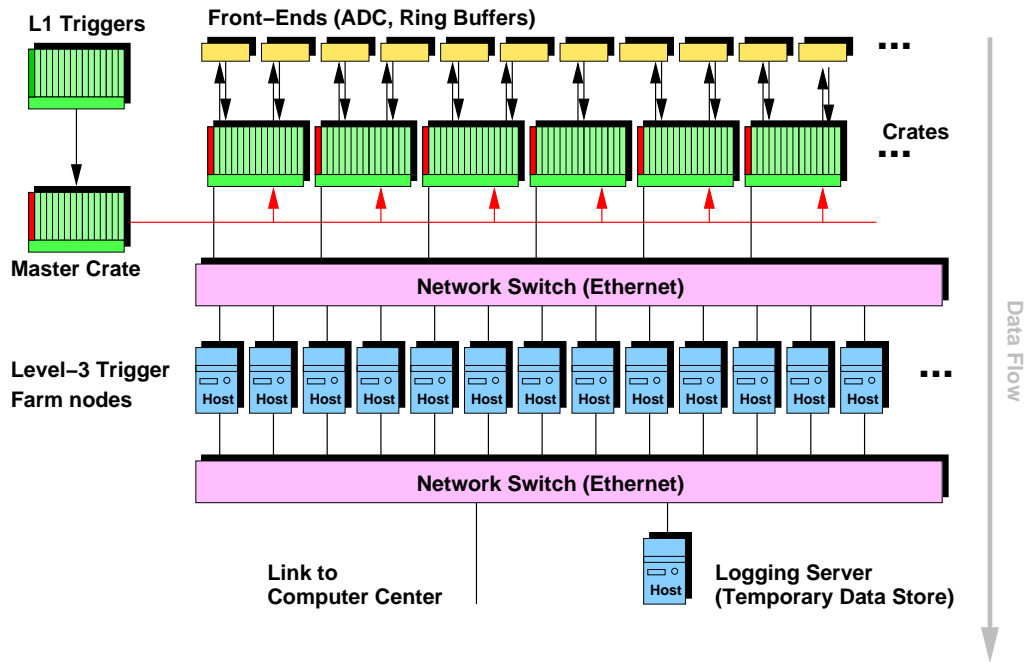}
\caption{Overview of the \babar{} Data Acquisition System}
\label{babardaq}
\end{figure*}

Figure \ref{babardaq} gives an overview of the \babar{} data
acquisition system. 

Sampling and analog-to-digital conversion takes place in the front end
electronics, mounted directly on or close to the subdetectors. From
there, data is sent to ``Read-Out-Modules'' (ROMs) which are 9u VME
modules that combine a commercial PowerPC single board computer
running the VxWorks operating system with custom-built components. The
number of VME crates and ROMs in the crate depends on design and data
load of the subdetector. The ROMs apply detector-specific feature
extraction algorithms and then transfer their event contributions via
the VME bus to the ROM in slot 0 of their crate. This is the first
stage of event building. The next stage is to combine all slot-0 ROM
contributions of an event by sending them to the same Level-3 trigger
farm node via Ethernet.  The network event builder uses a UDP-based
protocol, does not attempt to retransmit lost packets and employs a
simple window-based flow control mechanism. Events are sent to a
quasi-random sequence of farm nodes by using a hash value based on the
event time stamp.

In the farm node, the Level-3 filter algorithms are applied on the
complete event, events that pass the selection are sent to a central
logging server where they are written to disk files. These files are
subsequently transferred to the SLAC computer center for permanent tape
storage and near-real-time full reconstruction.

The initial design goal was 2kHz of 30kByte events to be read out with
negligible dead time from the front end electronics, feature
extracted, built and shipped to the Level-3 trigger that would select
100Hz of events for logging and permanent storage. With the original
farm of 32 330MHz Sun Ultra-5 machines the system did indeed reach
2kHz, limited by the amount of CPU time available to the Level-3
software trigger.

\section{PLANNING THE UPGRADE}

\subsection{Reasons for Upgrading}

A farm upgrade became necessary for the following reasons:

\begin{itemize}
\item The projected PEP-II luminosities and trigger rate
  extrapolations require the data acquisition system to handle higher
  rates and larger event sizes, in addition, operational experience
  showed that headroom of a factor of two is desirable to handle
  temporary high background conditions.
  
\item Having more CPU for more advanced Level-3 trigger algorithms
  will facilitate to keep the logging rate low.
  
\item The Sun Ultra-5 machines were reaching their end of life,
  showing an increasing number of hardware failures.
\end{itemize}

To be able to safely handle the projected trigger rates until at least 
2005, it was decided that an upgrade should increase the farm CPU
capacity by a factor of about 10 and that the 100MBit/s Ethernet connections 
in the network event builder should be upgraded to Gigabit Ethernet.

\subsection{Requirements and Upgrade Options}

Any upgrade option had to meet the following basic requirements:

\begin{itemize}
  
\item Run one of the two operating systems supported as \babar{}
  software platforms: RedHat Linux on i386 or Solaris/SPARC
  
\item Hardware must be available during the planning phase (fall 2001)
so that performance measurements could be obtained and to prevent
product availability from impacting the rigid deployment schedule
during the Summer 2002 shutdown.

\end{itemize}

Some options had been ruled out a priori, because they were expected
to be too different from the existing system and would have required
extensive R\&D and testing efforts, for example multiprocessor systems
with more than 2 CPUs where the concern were symmetric multiprocessing
scaling and interrupt distribution issues.

In the end, the following major options were considered:

\begin{enumerate}
  
\item \label{280R} Rack-mountable Sun 750MHz UltraSPARC-III dual-CPU
  systems (4 rack units high)
  
\item \label{netra} Replace Sun 330MHz Ultra-5 machines with
  rack-mountable 1-CPU 440MHz Sun machines (1 rack unit high), add
  systems as necessary and replace the farm later when faster and
  cheaper SPARC single or dual CPU systems became available
 
\item \label{intel} Intel 1.2Ghz Pentium-III Dual-CPU systems with
  Linux (1 rack unit high)
\end{enumerate}

Both Sun options (\ref{280R} and \ref{netra}) had the advantage that
no software modifications to ODF were required and that they could
just be used as a drop-in replacement. However, compared to the Intel
option (\ref{intel}), they were significantly more expensive and had
much larger rack space requirements: For comparable CPU power, option
\ref{280R} needed more rack space because of slower CPUs and a larger
form factor, option \ref{netra} just because of the much slower CPUs.

The Intel option had the major disadvantage to require modifications
to ODF and OEP software in order to accommodate the different byte
ordering. Option \ref{netra} had the additional disadvantage to be a
two-step upgrade where the second step relied on not yet existing
hardware.

\vspace{1em}

After extensive consideration of these options, the Intel Linux option
was chosen because it was expected to be the most cost-effective and
long-term maintainable solution.

\section{ADAPTING THE SOFTWARE}

\subsection{Data Flow Software}

ODF hardware and software is responsible for real-time readout, data
transport, feature extraction and multi-stage event building. The last
stage is a network event builder implemented on top of UDP, using no
retransmit and a simple end-to-end flow control protocol. The
underlying link-level protocol is Ethernet. To maximize system
performance, the original design required the machines on the
receiving end of the network event builder to have the same byte
ordering as the PowerPC embedded systems on the source side. For the
network event builder, the event data payload is transparent binary
data.

In order to accommodate Intel/Linux farm machines, ODF software had to
be modified. To avoid wasting precious CPU time on the embedded
systems, all byte swapping occurs on the farm machines. The ODF
network protocol was changed so that all datagram headers are 32-bit
aligned and swappable and whole datagrams (ODF headers as well as
event data payload) are 32-bit byte swapped after they have been
received by the farm machines. ODF software then builds the event and
delivers the complete pre-swapped event data to OEP.

Since the ODF network protocol always uses the source byte order, it
is independent of the byte order of the farm machines and allows the
use of Sun or Linux machines, in principle even in a mixed
farm. During testing we took advantage of this very important feature.

In parallel to upgrading the farm to Gigabit Ethernet, the embedded
systems were upgraded to Gigabit Ethernet. To minimize the overhead in
CPU and wall clock time for sending a packet, a custom minimal UDP
protocol stack and network device driver was developed. This driver
sends a 1500 byte packet in 17.3$\mu$s, compared to the 66$\mu$s of
the VxWorks driver that came with the Gigabit Ethernet interface card
and the 100$\mu$s needed by the 100MBit ethernet driver used before
the upgrade.

\subsection{Online Event Processing and Level-3 Filter Software}

OEP and Level-3 filter obviously need to look at the event data. While
all navigational information is 32-bit wide and aligned on 32-bit
boundaries, individual containers can hold 8-bit, 16-bit and 32-bit
quantities. As mentioned above, events are delivered to OEP
pre-swapped, that means that the navigational information is already
intact. The classes that describe the data format then know how to fix
up 8-bit and 16-bit quantities, this process only happens on
demand. The on-disk representation of data is always big-endian, so on
little endian machines the reverse process is applied before data is
written out to disk or sent over the network.

\subsection{System Software and System Management}

System management for Linux machines in the \babar{} online computing
infrastructure was achieved by modifying the \babar{}-online-specific
version of {\em Taylor} to support RedHat Linux \cite{redhat} and to
set up a {\em Kickstart} configuration for network installation of
Linux machines.  Taylor is a system post-install and configuration
language developed at SLAC \cite{taylor}.

\section{TESTING AND HARDWARE ACQUISITION}

\subsection{Choosing Hardware}

Since time and personnel resources for testing were limited, the
initial choice of hardware was made, considering experience from the
SLAC windows group who was already operating a number of Dell
PowerEdge 1550 1 rack-unit, 1.2GHz dual-CPU servers with Intel E1000
Gigabit Ethernet cards. We decided to use fiber-optic Gigabit Ethernet
because at decision time there was no significant experience at SLAC
with the cheaper copper cable Gigabit Ethernet. 

A small number of machines was acquired and used in different testing
modes:

\begin{itemize}

\item Teststand testing during software development was used to
develop the software, set up the system software infrastructure and
verify the basic suitability of the machines for the task.

\item ``Parasitic'' testing in the real system was used to find
problems with system stability or other operational issues. The
connection-less and byte order independent event builder protocol
allowed to use the port monitoring feature of the event building
switch to send real-time copies of datagrams to a small number of
machines being tested. In this mode, the switch was configured to send
copies of all packets addressed to a Sun farm node to the interface of
a Linux node that had been set up with the same MAC and IP addresses
as the original node but had been blocked from sending any replies
back.

To ensure proper operation, two machines were
run in this mode for several months, no problems were found. 

\item Small-scale testing of up to six new machines in the real
  system.

\end{itemize}

By the time of the purchase decision, the tested hardware was no
longer available and we had to go through another testing cycle with
the successor model (Dell PowerEdge 1650). At that time it was decided
to use 1.4GHz systems instead of 1.2GHz. 

Compared to the 330MHz UltraSPARC-II baseline, our application runs
2.75 times as fast on a 1.4GHz Pentium-III CPU. Because of limitations
in the number of Gigabit Ethernet switch ports and cost issues, we
decided to purchase 50 new machines in the final configuration:

\begin{itemize}
\item Dual-Pentium-III, 1.4 GHz
\item 1GByte RAM
\item optical Intel E1000 Gigabit Ethernet card
\item single power supply
\item no remote lights-out management capabilities (during data taking
there are always operators in the control room who can manually power
cycle a machine if necessary)
\end{itemize}

Compared to the old Sun farm the new farm provides {\em 8.5} times the 
CPU power.

\section{DEPLOYMENT AND RESULTS}

During a 4-month shutdown in summer 2002, the 50 new machines were
installed in 3 water-cooled racks in the \babar{} electronics hut.
Since the Dell rack rails could not easily be made fit into the
existing racks, pairs of machines were stacked on regular rack
shelves. The electronics hut is accessible during data taking, so it
was decided to save the cost of a remote power control or reset system
and simply install serial lines to the console ports. New modules with
a total of 80 Gigabit Ethernet ports were installed in the Cisco 6509
event building switch. To satisfy subdetector data acquisition needs
during the shutdown, a part of the old farm was kept in place and
operational. Installation and startup of the new farm was very smooth,
one machine was broken on arrival and another machine failed after one
week of operation.

\vspace{1em}

We did not encounter other major problems and since the start of
\babar{} Run 3 in November 2002 the new system is used in
production. After additional improvements to the feature extraction
code the system is now capable of handling {\em 5.5kHz} Level-1
trigger rate at 30kByte event size and nominal background rates, limited
by the feature extraction code in the ROMs. 

We did not observe any stability problems with the Linux farm
machines.

\vspace{1em}

\section{CONCLUSIONS}

The upgrade of the \babar{} data acquisition system to Linux and
Gigabit Ethernet has been successful and improved the maximum
rate capability at nominal backgrounds to 5.5kHz.

\section{ACKNOWLEDGEMENTS}

We would like to thank everyone who contributed to the upgrade project
and helped to actually make it happen on time.

%\bibliography{linuxupgrade}

\end{document}